\documentclass[aps,amssymb,amsmath,twocolumn]{revtex4} 
\usepackage{epsfig}
\usepackage[dvips]{color}

\begin{document}

\newcommand{\eins}{\mbox{$1 \hspace{-1.0mm}  {\bf l}$}}
\newcommand{\D}[3]{D_{#1:#2}(#3)}
\newcommand{\Done}[2]{D_{#1}(#2)}
\newcommand{\s}[2]{\sigma_{#1}^{(#2)}}
\newcommand{\im}{{\rm i}}
\newcommand{\Clus}{{\cal C}}
\newcommand{\half}{{\frac{1}{2}}}
\newcommand{\ket}[1]{|#1\rangle}
\newcommand{\bra}[1]{\langle#1|}
\newcommand{\braket}[2]{\langle#1|#2\rangle}
\newcommand{\vek}[1]{{\boldsymbol {#1}}}
\newcommand{\iden}{\eins\,}

\newcommand{\ncd}{\newcommand}
\ncd{\x}{$\bullet\,\,\,\,$}
\ncd{\oo}{$\mbox{}\,\,\,\,\,\,\,$}
\ncd{\nil}{$\bigcirc$}
\ncd{\pu}{$\bullet$}
\ncd{\ua}{$\uparrow$}
\ncd{\ra}{$\rightarrow$}
\ncd{\ds}{\displaystyle}
\ncd{\dummy}{\mbox{\tiny{\textcolor[cmyk]{0.5,0,0,0}{+}}}}
\ncd{\CNOT}{\mbox{CNOT}}
\ncd{\QC}{$\mbox{QC}_{\cal{C}}\,\,$}
\ncd{\QCns}{$\mbox{QC}_{\cal{C}}$}
\ncd{\fc}{\mbox{fc}}
\ncd{\bc}{\mbox{bc}}
\ncd{\notexists}{\exists \!\!\hspace*{-0.28mm}|\,\,\hspace*{0.28mm}}

\title{Computational Model for the One-Way Quantum Computer:\\
       Concepts and Summary}

\author{Robert Raussendorf, and Hans J. Briegel}
\affiliation{Sektion Physik, Ludwig-Maximilians-Universit{\"a}t M\"unchen,
Germany} 
  
\begin{abstract}
   The one-way quantum computer (\QCns) is a universal scheme of
   quantum computation consisting only of
   one-qubit measurements on a particular entangled multi-qubit state,
   the cluster state. The computational model underlying the \QC
   is different from the quantum logic network model and it is
   based on different constituents. It has no quantum register and does not
   consist of quantum gates. The \QC is nevertheless quantum
   mechanical since it uses a highly entangled cluster state as the
   central physical resource. The scheme works by measuring quantum
   correlations of the universal cluster state. 
\end{abstract}

\maketitle

\section{Introduction}
\label{intro}

Quantum computational models play a twofold role in the
development of quantum information science.
On the theoretical side, they provide the framework in which
mathematical concepts such as a ``computation''
or an ``algorithmic procedure'' become connected
to the laws of physics. Basic notions of computer science
such as ``computational complexity'' or `logical depth''
are usually derived with reference to such a model.
On the practical side, computational models can have
a strong influence on the design of actual experiments
that try to realize a quantum computer in the laboratory.

The first model of a quantum computer, the quantum Turing
machine (QTM) introduced by Deutsch \cite{QTM} and further developed by
Bernstein and Vazirani \cite{BV}, connects a
computation to a unitary transformation on the Hilbert space
spanned by all possible ``configurations'' of the machine.
Unlike its classical analog, it can be in a coherent
superposition of many different configurations at the same time,
which allows for the interference of different computational paths
during a computation. This distinct  feature of the QTM has opened
the room for the invention of  more efficient (quantum)
algorithms that make use of interference effects.

On the other hand, most proposals for implementing a quantum
computer in real physical systems do not follow the model
of a quantum Turing machine. The design of most of todays
experiments follow instead the model of a quantum logic network (QLN)
\cite{QLNW},\cite{Yao}. Although it was shown to be computationally
equivalent to the QTM \cite{QLNW},\cite{Yao}, this model has been
used more commonly in both theoretical and experimental
investigations. The notion of quantum gates makes it much simpler
to formulate quantum algorithms in the network language and most of the
quantum algorithms that one knows of today --including Shor's
celebrated factoring algorithm \cite{fac}-- have been
formulated within the network model. Furthermore, the fact that  
universal sets of quantum gates can  be realized from only two-qubit
interactions \cite{barenco95} has considerably simplified the problem
of identifying 
specific physical systems that are suitable \cite{DiV} for quantum
computation.  

In both the QTM and the QLN model of a quantum computer,
unitary evolution plays a key role, even though the way how such
a unitary evolution is generated is quite different. Recently, it has
become clear that quantum gates (and thus general unitary transformations)
need not be generated from a coherent Hamiltonian dynamics. Instead
several schemes \cite{QFT}-\cite{Nil} have been  proposed in which
projective von Neumann measurements play a constitutive role.

Recently, we introduced the scheme of the one-way quantum computer
\cite{QCmeas}. 
This scheme uses a given entangled state, the so-called cluster state
\cite{BR},
as its central physical resource. The entire quantum computation consists
only of a sequence one-qubit projective measurements on this entangled
state.
We called this scheme the ``one-way quantum computer'' since the
entanglement
in the cluster state is destroyed by the one-qubit measurements and
therefore
it can only be used once. While it is possible to simulate any unitary
evolution with the one-way quantum computer, the computational model
of the \QC makes no reference to the concept of unitary evolution. A quantum
computation
corresponds, instead, to a sequence of simple projections in the Hilbert
space
of the cluster state. The information that is  processed is extracted
from the measurement
outcomes and is thus a purely classical quantity.

As we have shown in \cite{QCmeas}, any quantum logic network
can be simulated efficiently on the one-way quantum
computer. This shows that the one-way quantum computer is, in fact,
universal. Surprisingly, it turns out that for many algorithms the
simulation of a unitary network can be parallelized to a higher degree than
the  original network itself. As an example, circuits in the
Clifford group --which is generated by the  CNOT-gates,
Hadamard-gates and $\pi/2$-phase shifts-- can be performed
by a \QC in a single time step, i.e. all the measurements to implement
such a circuit can be carried out at the same time. More generally, in
a simulation  
of a quantum logic network by a one-way quantum computer, the temporal
ordering of the gates of the network is transformed into a spatial
pattern of measurement bases for the individual qubits on the resource cluster
state. For the temporal ordering of the measurements there is, however, no
counterpart in the network model. Therefore, the question of
complexity of a quantum computation must be possibly revisited.
 
In the following we would like to give an introduction to the
computational
model that describes information processing with the
one-way quantum computer.
To stress the importance of the cluster state for the scheme, we will
use the abbreviation QC$_{\cal C}$ for ``one-way quantum computer''.
The computational model underlying the QC$_{\cal C}$ has been described
in a technical report in Ref.~\cite{model}. The purpose of the present
paper is to give a summary of this model, concentrating on the
concepts that we have introduced to
describe computation with the \QCns. We describe the objects that
comprise the information processed with the \QC and the temporal
structure of this processing. The reader who is interested in the
details of the derivations is referred to \cite{model}. 

\section{The \QC as a universal simulator of quantum logic networks}
\label{summary}

In this section, we give an outline of the universality proof
\cite{QCmeas} for the \QCns. To demonstrate universality we show that
the \QC can simulate any quantum logic network efficiently. It shall be pointed
out from the beginning that the network model does not provide the
most suitable description for the \QCns. Nevertheless, the
network model is the most widely used form of describing a quantum
computer and therefore the relation between the network model and
the \QC must be clarified. 

For the one-way quantum computer, the entire resource for the quantum 
computation is provided  initially in the form of a specific entangled
state --the cluster state \cite{BR}-- 
of a large number of qubits. Information is then written onto the 
cluster, processed, and read out from the cluster by one-particle 
measurements only. The entangled state of the cluster thereby serves as a 
universal ``substrate'' for any quantum computation. Cluster states
can be created  
efficiently in any system with a quantum Ising-type interaction (at very low 
temperatures) between two-state particles in a lattice
configuration. More specifically, to create a cluster state
$|\phi\rangle_{\cal{C}}$, the qubits on a cluster ${\cal{C}}$ 
are at first all prepared individually in a state $|+\rangle = 1/\sqrt{2}
(|0\rangle + |1\rangle)$ and then brought into a cluster state by
switching on the Ising-type interaction $H_{\mbox{\tiny{int}}}$ for an
appropriately chosen finite time span $T$. The time evolution operator
generated by this Hamiltonian which takes the initial
product state to the cluster state is denoted by $S$.

The quantum state $|\phi\rangle_{\cal{C}}$, the cluster state of a 
cluster $\cal C$ of neighbouring qubits, provides in advance all
entanglement that is involved in the subsequent quantum  
computation. It has been shown \cite{BR} 
that the cluster state $|\phi\rangle_{\cal C}$ is characterized by a
set of eigenvalue  equations
\begin{equation}
    \sigma_x^{(a)} \bigotimes_{a'\in ngbh(a)}\sigma_z^{(a')} 
    |\phi\rangle_{\cal C} = {(-1})^{\kappa_i} |\phi\rangle_{\cal C}, 
\label{EVeqn}
\end{equation} 
where $ngbh(a)$ specifies the sites of all qubits 
that interact with the qubit at site $a\in {\cal C}$ and $\kappa_i \in
\{0,1\}$ for all $i \in {\cal{C}}$. The equations (\ref{EVeqn}) are 
central for the proposed computation scheme. Cluster states specified
by different sets $\{\kappa_i,i \in {\cal{C}}\}$ are local unitary
equivalent, i.e. can be transformed into each other by local unitary
rotations of single qubits, and are thus equally good for
computation. In the following we 
will therefore confine ourselves to the case of 
\begin{equation}
    \kappa_i=0\,\; \forall i \in {\cal{C}}.
\end{equation}
 
It is important to realize
here that information processing is possible even though the result of
every measurement in any direction of the Bloch sphere is completely
random. The reason for the randomness of the measurement results is
that the reduced density operator for 
each qubit in the cluster state is $\frac{1}{2}{\bf{1}}$. While the
individual measurement results are irrelevant for the computation, the
strict correlations between measurement results inferred from 
(\ref{EVeqn}) are what makes the processing of quantum information
 on the \QC possible.

For clarity, let us emphasize that in the scheme of the \QC  we
distinguish between cluster qubits on 
${\cal{C}}$  which are measured in the process of computation, and the
logical qubits. The  
logical qubits constitute the quantum information being processed while
the cluster qubits in the initial cluster state form an entanglement resource.
Measurements of their individual one-qubit state drive the
computation.

\begin{figure}[tph]
\begin{center}
 \epsfig{file=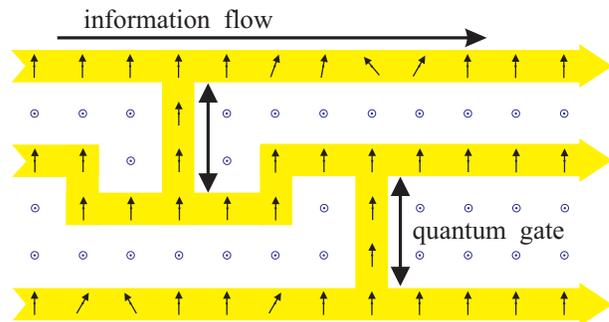,width=8cm}
 \caption{\label{FIGetching}Simulation of a
 quantum logic network by measuring two-state particles on a 
  lattice. Before the measurements the qubits are in the
  cluster state $|\phi\rangle_{\cal{C}}$ of (\ref{EVeqn}). 
  Circles $\odot$ symbolize measurements of $\sigma_z$, vertical arrows
  are measurements of $\sigma_x$, while tilted arrows refer to
  measurements in the x-y-plane.}
\end{center}
\end{figure}

To process quantum information with this cluster, it suffices to
measure its particles in a certain order and in a certain basis, as
depicted in Fig.~\ref{FIGetching}. Quantum 
information is thereby propagated through the cluster and
processed. Measurements of $\sigma_z$-observables effectively remove the
respective lattice qubit from the cluster. Measurements in the
$\sigma_x$- (and $\sigma_y$-) eigenbasis are
used for ``wires'', i.e. to propagate logical quantum bits through the
cluster, and for the CNOT-gate between two logical qubits. Observables
of the form $\cos(\varphi)\,\sigma_x \pm \sin(\varphi)\, \sigma_y$ are
measured to realize arbitrary rotations of logical qubits. 
For these cluster qubits, the basis in  which each of
them is  measured depends on the results of preceding 
measurements. This introduces a temporal order in which the
measurements have to be performed. The processing is finished once all qubits 
except a last one on each wire have been measured. The remaining
unmeasured qubits form the quantum register which is now ready to be
read out. At this point, the
results of previous measurements  
determine in which basis these ``output'' qubits need to be measured for the 
final readout, or if the readout measurements are in the $\sigma_x$-,
$\sigma_y$- 
or $\sigma_z$-eigenbasis, how the readout measurements have to be
interpreted. Without loss of generality, we assume in this paper that
the readout measurements are performed in the $\sigma_z$-eigenbasis. 

To understand the \QC in network model terms, in the
same way as we decompose networks into gates, we would like to decompose a
\QCns-circuit as a simulator of a quantum logic network into
simulations of quantum gates. This requires some adaption. First of
all, we need to identify a quantum input and -output. To do so, we
first modify the \QCns-computation slightly and later remove this
modification again. The modification is this: instead of creating a
universal cluster
state and subsequently measuring it we now allow for read-in of an
arbitrary quantum input. Then, the modified procedure consists of the
following steps. 1) Prepare a state
$|\psi_{\mbox{\footnotesize{in}}}\rangle_I \otimes (\bigotimes_{j \in
  {\cal{C}} \backslash I} |+\rangle_j)$ where
$|\psi_{\mbox{\footnotesize{in}}}\rangle$ is the input state prepared
on a subset of the cluster qubits $I \subset {\cal{C}}$. 2) Entangle
the state via the unitary evolution $S$ generated by the Ising
interaction. 3a) Measure all cluster qubits except for those of the
output register $O \subset {\cal{C}}$. In this way, the state
$|\psi_{\mbox{\footnotesize{in}}}\rangle$ is teleported from $I$ to
$O$ and at the same time processed. 3b) Measure the qubits in the
output register $O$ (readout). 

Please note that for the ``default input''
$|\psi_{\mbox{\footnotesize{in}}}\rangle_I = 
\bigotimes_{i \in I} |+\rangle_i$ this modified procedure is
equivalent to the original one. Then, steps 1 and 2 create a cluster
state (which could as well be created by any other method) and steps
3a) and 3b) form the sequence of measurements. As will be discussed in
Section~\ref{NoqIO}, as long as the quantum input is known it is
sufficient to consider $\bigotimes_{i \in I} |+\rangle_i$ and thus
one-qubit measurements on cluster states.

Steps 1) - 3a) form the procedure to simulate some unitary
network applied to the quantum register. It is decomposed into
similar sub-procedures for gate simulation:
loading the input, entangling operation, measurement of all but the
output qubits. Provided the measurement bases have been chosen
appropriately, a procedure of this type teleports a general input
state from one part of the cluster to another and thereby also
processes it. 

We explain the \QC as a succession of gate simulations, i.e. as
repeated steps of entangling operations and measurements on
sub-clusters. In reality, however, a different scheme is realized, namely first
all the cluster qubits are entangled and second they are measured. These
to ways to proceed are mathematically equivalent as has been
demonstrated in \cite{QCmeas}. The basic reason for this equivalence
is that, in the sequential picture, later entangling operations
commute with earlier measurements because they act on different
particles. Therefore, the order of operations can be interchanged such
that first all entangling operations and after that all measurements
are performed.

\begin{figure}
    \begin{center}
        \begin{tabular}{cc}
            \multicolumn{2}{c}{\parbox{5.7cm}{\begin{center}\epsfig{
            file=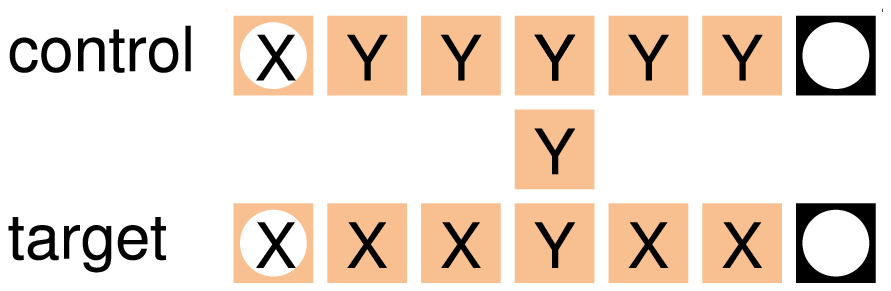, width=5.7cm} \\ CNOT-gate \end{center}}} 
            \\
            \parbox{3.2cm}{\epsfig{file=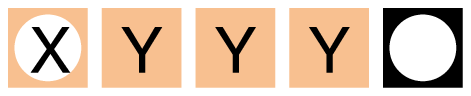,width=3.2cm}} &
            \parbox{3.2cm}{\epsfig{file=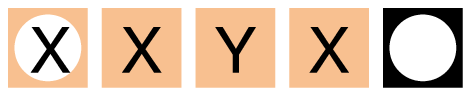,width=3.2cm}}
            \vspace{0.15cm}\\ 
             Hadamard-gate & $\pi/2$-phase gate \\
             \multicolumn{2}{c}{\parbox{3.2cm}{\begin{center}\epsfig{
             file=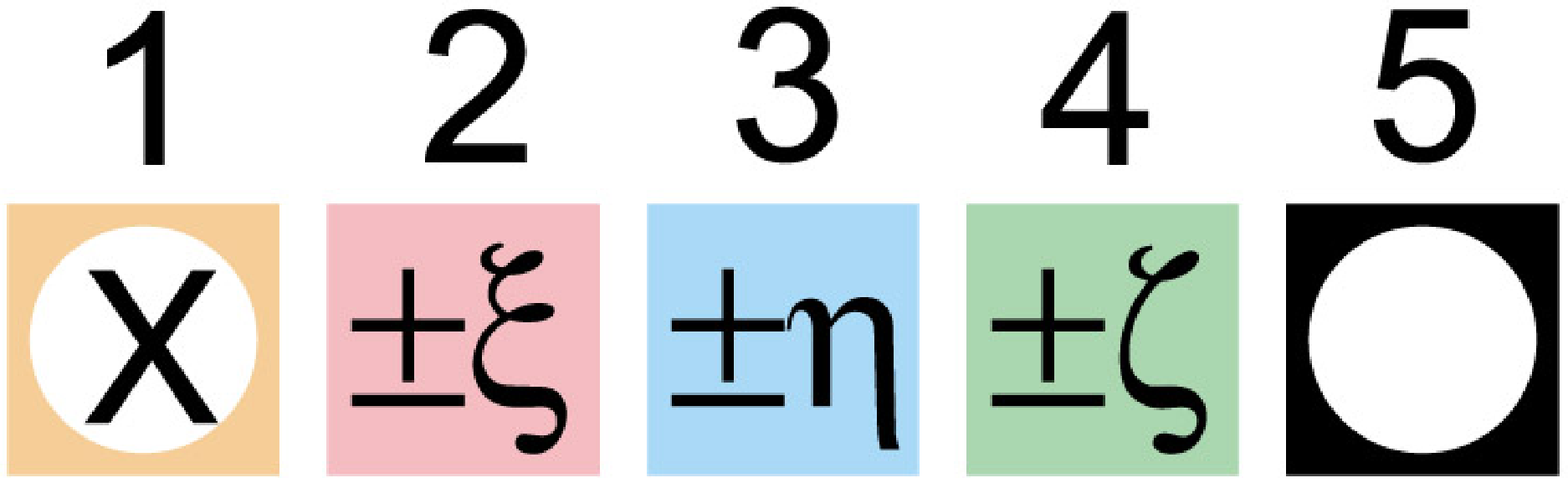, width=3.2cm} \\ rotation
            \end{center}}}  
        \end{tabular}
        
        \caption{\label{Gates}{Realization of 
              the required gates on the \QCns.  CNOT-gate between
              neighbouring qubits, the Hadamard gate, the $\pi/2$
              phase gate and the general rotation specified by the
              three Euler angles $\xi$, $\eta$, $\zeta$.}}
    \end{center}
\end{figure}

In the following we review two points of the
universality proof for the \QCns: 
the realization of the arbitrary one-qubit rotation as a member of the
universal set of gates, and the effect of the randomness of
the individual measurement results and how to account for them. For
the realization of a CNOT-gate see Fig.~\ref{Gates} and \cite{QCmeas}.

An arbitrary rotation $U_R \in SU(2)$ can be achieved in a
chain of 5 qubits. Consider a
rotation in its Euler representation 
\begin{equation}
    \label{Euler}
    U_R(\xi,\eta,\zeta) = U_x(\zeta)U_z(\eta) U_x(\xi),
\end{equation}
where the rotations about the $x$- and $z$-axis are 
$
        U_x(\alpha) =
        \displaystyle{\mbox{exp}\left(-i\alpha\frac{\sigma_x}{2}\right)} 
$ and
$  
        U_z(\alpha) = \displaystyle{\mbox{exp}\left(-i\alpha 
        \frac{\sigma_z}{2}\right)} 
$.
Initially, the
first qubit is in some state
$|\psi_{\mbox{\footnotesize{in}}}\rangle$, which is to be rotated, and
the other qubits are in 
$|+\rangle$. After the 5 qubits are entangled by the time
evolution operator $S$ generated by the Ising-type Hamiltonian, the
state $|\psi_{\mbox{\footnotesize{in}}}\rangle$ can be rotated by measuring qubits 1 to
4. At the same time, the state is also transfered to site 5. The qubits $1
\dots 4$ are measured in appropriately  chosen bases, {\em{viz.}}
\begin{equation}
    \label{Measbas}
    {\cal{B}}_j(\varphi_{j}) = \left\{
        \frac{|0\rangle_j+e^{i \varphi_{j}} |1\rangle_j}{\sqrt{2}} ,\,
        \frac{|0\rangle_j-e^{i \varphi_{j}} |1\rangle_j}{\sqrt{2}}
    \right\}
\end{equation} 
whereby the measurement outcomes  $s_{j} \in \{ 0,1 \}$ for
$j=1\dots 4$ are obtained. Here, $s_{j}=0$ means that qubit $j$ is projected 
into the first state of
${\cal{B}}_j(\varphi_{j})$. In
 (\ref{Measbas}) the basis states of all possible measurement bases
lie on the equator of the Bloch sphere, i.e. on the intersection of
the Bloch sphere with the $x-y$-plane. Therefore, the measurement
basis for qubit $j$ can be specified by a single parameter, the
measurement angle $\varphi_{j}$. The
measurement direction of qubit $j$ is the vector on the Bloch sphere
which corresponds 
to the first state in the measurement basis
${\cal{B}}_j(\varphi_{j})$. Thus, the
measurement angle $\varphi_{j}$ is equal to the
angle between the measurement direction at qubit $j$ and the positive
$x$-axis. For all of
the gates constructed so far, the cluster qubits are either 
--if they are not required for the realization of the
circuit-- measured in $\sigma_z$, or --if they are required-- measured
in some measurement direction in the $x-y$-plane. 
In summary,
the procedure to implement an arbitrary rotation $U_R(\xi,\eta,\zeta)$, 
specified by its Euler angles $\xi,\eta,\zeta$, is this:
1. measure qubit 1 in ${\cal{B}}_1(0)$; 2. measure qubit 2 in
 ${\cal{B}}_2\left(-(-1)^{s_1} \xi \right)$; 3. measure qubit 3 in
 ${\cal{B}}_3\left(-(-1)^{s_2} \eta \right)$; 4. measure qubit 4 in 
 ${\cal{B}}_4\left(-(-1)^{s_1+s_3} \zeta \right)$. In this way the
rotation $U_R^\prime$ is realized:
\begin{equation}
    \label{Rotprime}
    U_R^\prime(\xi,\eta,\zeta) =  U_\Sigma U_R(\xi,\eta,\zeta).
\end{equation}
The random byproduct operator 
\begin{equation}
    \label{Byprod1}
    U_\Sigma= \sigma_x^{s_2+s_4}\sigma_z^{s_1+s_3}
\end{equation} can
be corrected for at the end of the computation, as explained next.  

The randomness of the measurement results does not
jeopardize the function of the  circuit. Depending on
the measurement results, extra rotations $\sigma_x$ and $\sigma_z$ act on
the output qubits of every implemented gate, as in (\ref{Rotprime}), for
example. By use of the propagation
relations
\begin{equation}
    \label{Rotprop}
    \begin{array}{l}
    U_R(\xi,\eta,\zeta) \, \sigma_z^{s} \sigma_x^{s'} =\\  
    \hspace*{0.7cm}\sigma_z^{s} \sigma_x^{s'} \,
    U_R((-1)^{s}\xi,(-1)^{s'}\eta,(-1)^{s}\zeta),
    \end{array}
\end{equation}
\begin{equation} 
    \label{CNTprop}
    \begin{array}{l}
    \mbox{CNOT} (c,t) \, {\sigma_z^{(t)}}^{s_t} {\sigma_z^{(c)}}^{s_c}
    {\sigma_x^{(t)}}^{s_t'} {\sigma_x^{(c)}}^{s_c'} 
    =\\
    \hspace*{0.7cm}{\sigma_z^{(t)}}^{s_t} {\sigma_z^{(c)}}^{s_c+s_t}
    {\sigma_x^{(t)}}^{s_c'+s_t'} {\sigma_x^{(c)}}^{s_c'} \, 
    \mbox{CNOT}(c,t) ,
    \end{array}
\end{equation}  
these extra rotations can be pulled
through the network to act upon the output state. There they can be
accounted for by properly interpreting the $\sigma_z$-readout
measurement results.

The propagation relations for the general rotations and for the CNOT
gate, respectively, are different in the following respect. In the
propagation relation for the CNOT gate the gate remains unchanged and
the byproduct operator is modified. For rotations this is in general not
possible if one demands that the byproduct operator must remain in the
Pauli group, which is essential. Therefore, in the 
propagation relations for rotations the
gate changes and the byproduct operator remains unmodified. This is
the origin of adaptive measurement bases and thus the temporal
structure of \QCns-algorithms.

The reason why the propagation relation for the CNOT gate takes the
form (\ref{CNTprop}) is that it is in the Clifford group,
the normalizer of the Pauli group, which means that a Pauli operator is mapped
onto a Pauli operator under conjugation with any Clifford group
element. There are also local rotations in the Clifford group, among
them the Hadamard transformation and the $\pi/2$-phase gate. These
rotations can be simulated more efficiently than by the procedure for
general rotations described above. To see this, note that the
measurement bases ${\cal{B}}(\varphi)$ and ${\cal{B}}(-\varphi)$ 
in (\ref{Measbas})
coincide for angles $\varphi=0$ and for $\varphi=\pm\pi/2$. For
$\varphi=0$ the measurement basis ${\cal{B}}(\varphi)$ is the
eigenbasis of $\sigma_x$, and 
for $\varphi=\pm\pi/2$ the measurement basis ${\cal{B}}(\varphi)$ is
the eigenbasis of 
$\sigma_y$. In these cases, the choice of the measurement basis is not
influenced by the results of measurements at other qubits.  
The Hadamard gate and
the $\pi/2$ phase shift are such rotations. As displayed in
Fig.~\ref{Gates}, they are both realized by performing a
pattern of $\sigma_x$- and  $\sigma_y$-measurements on the cluster
${\cal{C}}$. The byproduct operators which are thereby created are
\begin{equation}
    \label{Hadabyprop}
    \begin{array}{rcl}
          U_{\Sigma,H} &=&
          \sigma_x^{s_1+s_3+s_4}\, \sigma_z^{s_2+s_3}\\
          U_{\Sigma,U_z(\pi/2)} &=&
          \sigma_x^{s_2+s_4}\,
          \sigma_z^{s_1+s_2+s_3+1}.
    \end{array}      
\end{equation}
Owing to the fact that the Hadamard- and
the $\pi/2$-phase gate are in the Clifford group, the propagation
relations for these rotations 
can also be written in a form resembling the
propagation relation (\ref{CNTprop}) for the CNOT-gate 
\begin{equation}
    \label{Hadaprop}
    \begin{array}{rcl}
        H\,{\sigma_x}^{s_x}{\sigma_z}^{s_z} &=&
        {\sigma_x}^{s_z}{\sigma_z}^{s_x}\, H, \\
        U_z(\pi/2)\,{\sigma_x}^{s_x}{\sigma_z}^{s_z} &=&
        {\sigma_x}^{s_x}{\sigma_z}^{s_x+s_z}\, U_z(\pi/2).
    \end{array}
\end{equation}

As stated above, the measurement bases to implement the Hadamard- and
the $\pi/2$-phase gate require no adjustment since only operators
$\sigma_x$ and $\sigma_y$ are measured. The same holds for the
implementation of the CNOT gate, see Fig.~\ref{Gates}. Thus, all the
Hadamard-, $\pi/2$-phase- and CNOT-gates of a quantum circuit can be
implemented simultaneously in the first measurement round with no
regard to their location in the network. In particular, quantum
circuits which consist only of such gates, i.e. circuits in the
Clifford group, can be realized in a single time step. As an example,
many circuits for coding and decoding are in the Clifford group.  

The fact that quantum circuits in the Clifford group can be realized
in a single time step has previously not been known for networks. The
best upper bound on the logical depth known so far scales
logarithmically with the number of logical qubits \cite{M&N}. 
One might therefore wonder whether the \QC is more efficient
than a quantum computer realized as a quantum logic network. This is
not the case in so far 
as both the quantum logic network computer and the \QC
can simulate each other efficiently. The fact that
each quantum logic network can be simulated on the \QC has been shown
in \cite{QCmeas}. The converse is also true because a resource cluster
state of arbitrary size can be created by a quantum logic network of
constant logical depth. Furthermore, the subsequent one-qubit measurements
are within the set of standard tools employed in the network
scheme of computation. In this sense, the operation of the \QC can be
cast entirely in network language.

However, while the network model comprises the
means that are used in a computation on a \QCns, it cannot describe
{\em{how}} they have to be used. In particular, in the above
construction --where a \QC simulating a quantum logic network is itself
simulated by a more complicated network-- the temporal order of the
measurements and the rules to adapt the measurement bases are not
provided with the network description. But without this additional information
the network to simulate the \QC is incomplete. 

It should be noted that a
link between the degree of parallelization of unitary operations and
the logical depth of a quantum algorithm does not exist a priori. It is
established only if quantum computation is identified with unitary
evolution.  The network model allows
statements about how much one can parallelize networks composed of
unitary gates. As an example, two unitary gates $U_1$, $U_2$ cannot be 
performed in parallel if they do not commute. For simulations of such
gates with the \QCns, however, this general restriction does not apply: The
simulations of two non-commuting gates can still
be parallelized if the gates are in the Clifford group.

With this observation we complete the survey of the universality proof
\cite{QCmeas} for the \QCns. To summarize, for simulation of a quantum
logic network on a one-way quantum computer, a set of universal gates can 
be realized by one-qubit measurements and the gates can be combined to
circuits. Due to the randomness of the results of the individual
measurements, extra byproduct operators occur. These byproduct
operators specify how the readout of the simulated quantum register
has to be interpreted. Also, they influence the bases of the one-qubit
measurements. 

In this section we have described the \QC as a simulator of quantum logic
networks. We adopted all the network notions such as the ``quantum
register'' and ``quantum 
gates''. We have found an additional structure, the byproduct operator,
which keeps track of the randomness introduced by the measurements.
In a network-like description of the \QCns, the byproduct operator
appears as some unwanted extra complication that has to be and
fortunately can be handled. In the next section we will point out
in which respect the description of the \QC as a network simulator is not 
adequate and in Section~\ref{model} we will present a different
computational model for the \QCns. For this model it will turn out
that the  byproduct operators form, in fact, the central quantities of
information processing with the \QCns, and that the ``quantum
register'' and the ``quantum gates'' disappear.     

\section{Non-network character of the \QC}
\label{NoqIO}

In the network model of quantum computation one usually regards the
state of a quantum register as the carrier of 
information. The quantum register is prepared in some input state and
processed to some output state by applying a suitable unitary
transformation composed of quantum gates. Finally, the output state of
the quantum register is measured by which the classical readout is
obtained. 

In this section we explain why the notions of ``quantum
input'' and ``quantum output'' have no genuine meaning for the \QC if we
restrict ourselves to the situation where the 
quantum input state is {\em{known}}. Shor's factoring
algorithm \cite{fac} and Grover's data base search algorithm
\cite{searoot} are both examples of such a situation. In these
algorithms one always
starts with the input state $\bigotimes_{i=1}^n
1/\sqrt{2}(|0 \rangle_i + |1 \rangle_i) = |++...+\rangle$. Other scenarios are
conceivable, e.g. where an unknown 
quantum input is processed and the classical result of the computation
is retransmitted to the sender of the input state; or the unmeasured
network output register state is retransmitted. These scenarios would lead
only to minor modifications in the computational model. How to
process an unknown quantum state has been briefly discussed in
Section~\ref{summary} but is not in the focus of this paper. So,
let us assume that the quantum input is known. There it is sufficient
to discuss the situation where an input state $|++..+\rangle_I$ is
read in on some subset $I \subset {\cal{C}}$ of the cluster
${\cal{C}}$. Any other known input state can be created on the cluster
from the standard quantum state $|++..+\rangle$, by some
circuit preceding the main one.

Reading in an input state $\bigotimes_{i \in I} |+\rangle_i$ means to
prepare the state $S [\bigotimes_{i \in I} |+\rangle_i \otimes
(\bigotimes_{j \in {\cal{C}}\backslash I} |+\rangle_j) ] =
|\phi\rangle_{\cal{C}}$, i.e. to prepare nothing but a cluster state.
The cluster state $|\phi\rangle_{\cal{C}}$ is a universal resource, no
input dependent information specifies it. In this sense, the \QC has
no quantum input. 

Similarly, the \QC has no quantum
output. Of course, the final result of any computation --including
quantum computations-- is a
classical number, but for the quantum logic network the state of the
output register before the readout measurements plays a distinguished
role. For the \QC this is not the case, there are just cluster qubits
measured in a certain order and basis. The measurement outcomes
contribute all to the result of the computation. 

We have identified subsets $I$, $O$ on the cluster ${\cal{C}}$ --$I$ for the
subset of the cluster which simulates the quantum register in its
input state and $O$ to simulate the quantum register in its output
state-- only to make the \QC suitable for a description in terms of
the network model. Such a terminology is not required for the \QC a
priori. It is not even appropriate: if, to perform a
particular algorithm on the \QCns, a quantum logic network is
implemented on a cluster state there is a subset of cluster qubits
which play the role of the output register. These qubits are not the
final ones to be measured, but among the first (!).  

As we have seen, the measurement outcomes from all 
the cluster qubits contribute to the result of the computation. The qubits
from $O \subset {\cal{C}}$ simulate the output state of the quantum
register and thus contribute directly. The cluster qubits in the set
$I \subset {\cal{C}}$ simulate the input state of the quantum register
and the outcomes obtained in their measurement contribute via the
accumulated byproduct operator that is required to interpret the
readout measurements on $O$. Finally, the qubits in the section $M
\subset {\cal{C}}$ of the cluster by whose measurements the quantum
gates are simulated also contribute via the byproduct operator. 

Naturally there arises the question whether there is any difference in
the way how
measurements of cluster qubits in $I$, $O$ or $M$ contribute to the
final result of the computation. As shown in \cite{model}, it turns out
that there is none. This is why we abandon the notions of quantum
input and quantum output altogether from the description of the \QCns.

\section{Computational model}
\label{model}

Quantum gates are not constitutive elements of the \QCns; these are instead
one-qubit measurements performed in a
certain temporal order and in a spatial pattern of adaptive
measurement bases. The most efficient temporal order of the
measurements does not 
follow from the temporal order of the simulated gates in the network
model. Therefore, a set of rules is required by which the optimal
order of measurements can be inferred. 
Generally, for circuits which involve a vast number
of measurements and subsequent conditional processing, it becomes
essential to have an additional structure to process the classical
information gained by the measurements. The \QC provides such a structure - the information flow vector
${\bf{I}}(t)$ (see \cite{model} and below). In fact, for the
computational model of the \QC this classical binary-valued quantity will turn
out to be the central object for information processing. 

So let us take a step back and look what the \QC is. On the quantum
level, the \QC works by measuring quantum correlations of the initial
universal cluster state. With the creation of the universal cluster state these
quantum correlations are provided before the computation starts.  
In contrast to the network model, they 
are not created in a procedure specific to the computational
problem. Therefore, for the \QC there is no processing of information
on the quantum 
level. In this sense, besides no quantum input and no quantum
output, the \QC has no quantum register either. 

Central from the conceptual point of view but also vital for the
practical realizability of the scheme is that the quantum correlations
of the cluster state can be measured qubit-wise. This requires a
temporal ordering of the measurements and adaptive 
measurement bases in accordance with previously obtained measurement
results. Thus there is processing at the classical level. 

\begin{figure}
    \begin{center}
        \epsfig{file=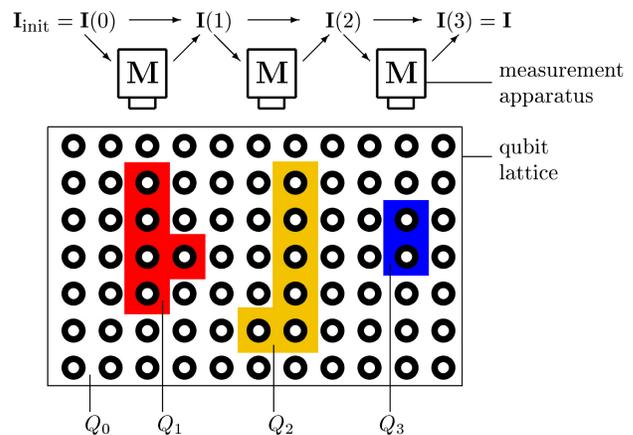,width=8.5cm}
        \caption{\label{scheme}{General scheme of
        the quantum computer via one-qubit measurements. The sets
    $Q_t$ of lattice qubits are measured one after the other. The
    results of earlier measurements determine the measurement bases of
    later ones. All classical information from the measurement results
    needed to steer the \QC is contained in the information flow vector
    ${\bf{I}}(t)$. After the last measurement round
    $t_{\mbox{\footnotesize{max}}}$,
    ${\bf{I}}(t_{\mbox{\footnotesize{max}}})$ contains the result of
    the computation.}}
    \end{center}
\end{figure}

The general view of a \QCns-computation is as follows. The cluster
${\cal{C}}$ is divided into disjoint subsets $Q_t \subset {\cal{C}}$
with $0 \leq t \leq t_{\mbox{\footnotesize{max}}}$, i.e.
$\bigcup_{t=0}^{t_{\mbox{\tiny{max}}}}Q_t = {\cal{C}}$ and $Q_s \cap
Q_t = \emptyset$ for all $s\not=t$. The cluster qubits within each set
$Q_t$ can be measured simultaneously and the sets are measured one
after another. The set $Q_0$ consists of all those qubits of which no
measurement bases have to be adjusted, i.e. those of which the
operator $\sigma_x$, $\sigma_y$ or $\sigma_z$ is measured. This
comprises all the redundant qubits, the qubits to implement the
Clifford part of the circuit and the qubits which simulate the network
quantum output. In the subsequent measurement rounds only operators of
the form $\cos \varphi \, \sigma_x +\pm \sin \varphi \, \sigma_y$ are
measured where $|\varphi|<\pi/2$, $\varphi \neq 0$. The
measurement bases are adaptive in these rounds. The measurement
outcomes from the qubits in $Q_0$ specify the measurement bases for
the qubits in $Q_1$ which are measured in the second round, those from
$Q_0$ and $Q_1$ together specify the bases for the measurements of the
qubits in $Q_2$ which are measured in the third round, and so
on. Finally, the result of the computation is 
calculated from the measurement outcomes in all the measurement rounds.     

Now there arise two questions. First, ``Given a quantum algorithm, how
can one find the measurement pattern and in particular the temporal
order in which the measurements are performed?''. As for the
measurement pattern, apart from a few exceptions such as the quantum
Fourier transformation or the quantum adding circuit \cite{LongMeas}
presently we know no better than to straightforwardly simulate the
network. Even 
then the optimal temporal order of the measurements is, as stated before,
different from what one expects from the order of gates in the
quantum logic network. The discussion of temporal complexity within the 
\QC will lead us to objects such as the {\em{forward cones}}, the
{\em{byproduct images}} and the {\em{information flow vector}}
\cite{model} which will be briefly introduced below. The second
question is: ``How complicated is the required classical
processing?''. In principle it could be that all the obtained
measurement results had to be stored separately and the functions to
compute the measurement bases were so complicated that one would gain
no advantage over the classical algorithm for the considered problem.
This is not at all the case. If the network algorithm runs on
$n$ qubits then the classical data that the \QC has to keep track of
is all contained in a
$2n$-component binary valued vector, the information flow vector
${\bf{I}}(t)$. The update of ${\bf{I}}(t)$, the calculation to adapt the 
measurement bases of cluster qubits according to previous measurement outcomes
and the final identification of the computational result are all elementary.
 
\begin{figure}[tph]
\begin{center}
 \setlength{\unitlength}{0.6cm}
\definecolor{lgray}{gray}{0.35}
\begin{picture}(12,6)
  \color{lgray}
  \newcounter{ic}
  \setcounter{ic}{-1}
  \setlength{\unitlength}{0.3mm}
  \multiput(120,70)(1.5,-0.5){80}{\stepcounter{ic}\line(0,1)\theic}
   \setcounter{ic}{-1}
  \multiput(120,70)(-1.5,-0.5){80}{\stepcounter{ic}\line(0,1)\theic}
  \setlength{\unitlength}{0.6cm}\color{white}
  \put(2.5,3.5){\circle*{0.77}}
  \put(7.5,3.5){\circle*{0.77}}
  \put(9.5,3.5){\circle*{0.77}}
  \put(8.5,5.5){\circle*{0.77}}
  \color{black}
  \setlength{\linewidth}{0.1mm}
  \put(0,1.5){\line(3,1){12}}
  \put(0,5.5){\line(3,-1){12}}
  \setlength{\linewidth}{0.7mm}
  \put(0,5.5){\line(1,0){12}}
  \put(0,4.5){\line(1,0){12}}
  \put(0,3.5){\line(1,0){12}}
  \put(0,2.5){\line(1,0){12}}
  \put(0,1.5){\line(1,0){12}}

  \put(2.5,4.5){\line(0,-1){1.37}}
  \put(7.5,4.5){\line(0,-1){1.37}}
  \put(9.5,2.5){\line(0,1){1.37}}
  \put(8.5,4.5){\line(0,1){1.37}}
  \put(2.5,3.5){\circle{0.77}}
  \put(7.5,3.5){\circle{0.77}}
  \put(9.5,3.5){\circle{0.77}}
  \put(8.5,5.5){\circle{0.77}}
  \put(2.5,4.5){\circle*{0.2}}
  \put(7.5,4.5){\circle*{0.2}}
  \put(9.5,2.5){\circle*{0.2}}
  \put(8.5,4.5){\circle*{0.2}}
  \put(6.0,3.5){\circle*{0.2}}

  \put(0,5.5){\circle*{0.35}}
  \put(0,4.5){\circle*{0.35}}
  \put(0,3.5){\circle*{0.35}}
  \put(0,2.5){\circle*{0.35}}
  \put(0,1.5){\circle*{0.35}}

  \put(12,5.5){\circle*{0.35}}
  \put(12,4.5){\circle*{0.35}}
  \put(12,3.5){\circle*{0.35}}
  \put(12,2.5){\circle*{0.35}}
  \put(12,1.5){\circle*{0.35}}
  \color{white}
  \put(0,5.5){\circle*{0.25}}
  \put(0,4.5){\circle*{0.25}}
  \put(0,3.5){\circle*{0.25}}
  \put(0,2.5){\circle*{0.25}}
  \put(0,1.5){\circle*{0.25}}

  \put(12,5.5){\circle*{0.25}}
  \put(12,4.5){\circle*{0.25}}
  \put(12,3.5){\circle*{0.25}}
  \put(12,2.5){\circle*{0.25}}
  \put(12,1.5){\circle*{0.25}}
  \multiput(8.1,3.1)(0,0.016){50}{\line(1,0){0.8}}
  \multiput(10.4,2.1)(0,0.016){50}{\line(1,0){0.8}}
  \color{black}
  \multiput(10.4,3.1)(0,0.016){50}{\line(1,0){0.8}}
  \multiput(10.4,4.1)(0,0.016){50}{\line(1,0){0.8}}
  \multiput(0.8,3.1)(0,0.016){50}{\line(1,0){0.8}}
  \multiput(0.8,4.1)(0,0.016){50}{\line(1,0){0.8}}
  
  \put(8.1,3.1){\line(1,0){0.82}}
  \put(8.1,3.9){\line(1,0){0.82}}
  \put(8.1,3.1){\line(0,1){0.8}}
  \put(8.9,3.1){\line(0,1){0.8}}

  \put(10.4,2.1){\line(1,0){0.81}}
  \put(10.4,2.9){\line(1,0){0.81}}
  \put(10.4,2.1){\line(0,1){0.8}}
  \put(11.2,2.1){\line(0,1){0.8}}

  \put(5.8,3.7){\normalsize{$\sigma_z$}}
  \put(5.8,2.9){\normalsize{$k$}}
  \put(1.6,4.0){\normalsize{$i$}}
  \put(1.6,3.0){\normalsize{$j$}}
  \put(8.95,2.9){\normalsize{$l$}}
  \put(11.3,4.1){\normalsize{$m$}}
  \put(11.3,3.1){\normalsize{$n$}}
  \put(11.3,2.1){\normalsize{$o$}}
  
  \put(10.55,2.4){\small{$U_z$}}
  \put(8.3,3.4){\small{$U_z$}}
  \color{white}
  \put(10.55,3.4){\small{$U_x$}}
  \put(10.55,4.4){\small{$U_x$}}
  \put(0.9,3.4){\small{$U_x$}}
  \put(0.9,4.4){\small{$U_x$}}
  \color{black}
  \put(12,1.2){\line(0,1){4.6}}
  \put(11.4,0.5){$U_{\Sigma,k}|_\Omega$}
  \setlength{\linewidth}{0.1mm}
  
  \put(-0.3,5.9){IN}
  \put(11.5,5.9){OUT}

\end{picture}
 \caption{\label{fc/bc}Forward and backward
 cones. The measurement of a cluster qubit $k$ for the implementation
 of the shown quantum logic network may, depending on the measurement
 outcome, result in a byproduct operator $\sigma_z$ (the underlying
 cluster and measurement pattern is not shown). This byproduct
 operator is propagated forward to act upon the output register as
 $U_{\Sigma,k}|_\Omega$. In forward propagation, it flips the
 measurement angles of the cluster qubits $m$, $n$ by whose
 measurement one-qubit rotations are implemented. The cluster qubits $m$ and
 $n$ are thus in the forward cone of $k$, $m,n \in \mbox{fc}(k)$, while $l$
 and $o$ are not. Similarly, $i,j \in \mbox{bc}(k)$.}
\end{center}
\end{figure}
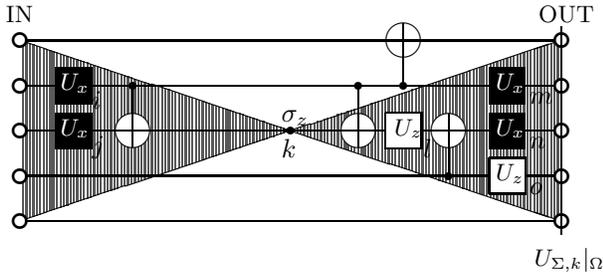

Let us first discuss the temporal ordering of the measurements. To
understand how the sets $Q_t$ of simultaneously measurable qubits are
constructed we introduce the notion of forward cones. The forward cone
$\mbox{fc}(k)$ of a cluster qubit $k \in 
{\cal{C}}$ is the set of all those cluster qubits $j \in {\cal{C}}$
whose measurement basis
${\cal{B}}(\varphi_{j,\mbox{\footnotesize{meas}}})$ 
depends on the result $s_k$ of the measurement
of qubit $k$ after the byproduct operator
${\left(U_k\right)}^{s_k}$ is propagated forward from the output side of the
gate for whose implementation the cluster qubit was measured
to the output side of the network. See Fig.~\ref{fc/bc}. Similarly, the 
backward cone $\mbox{bc}(k)$ of a cluster qubit $k \in
{\cal{C}}$ whose measurement bases depend upon the measurement result
at qubit $k$ when the byproduct
operator is propagated backward to the input side of the
network. The method to calculate the forward and backward cones
follows immediately from their definitions. Quite surprisingly, it
will turn out that only backward 
cones will appear in the computational model that finally
emerges. Nevertheless, the forward cones are used to identify the sets
of simultaneously measurable qubits as is explained below. 

What does it mean that a cluster qubit $j$ is in the forward cone of
another cluster qubit $k$, $j\in \fc(k)$? According to the definition,
a byproduct operator created via the measurement at cluster qubit $k$
influences the measurement angle
$\varphi_{j,\mbox{\footnotesize{meas}}}$ at cluster qubit $j$. To
determine the measurement angle at $j$ 
one must thus wait for the measurement result at $k$.  Therefore, the
forward cones generate a temporal ordering among the 
measurements. If $j \in
\fc(k)$, the measurement at qubit $j$ is performed later than that at
qubit $k$. This we denote by $k \prec j$
\begin{equation}
    \label{coneImpl}
    j \in \fc(k) \Rightarrow k \prec j.
\end{equation}  
The relation ``$\prec$'' is a strict partial ordering, i.e. it is
transitive and anti-reflexive. Anti-reflexivity is required for the scheme
to be deterministic. Transitivity we use to generate ``$\prec$'' from the
forward cones. This partial ordering can now be used
to construct the sets  $Q_t \subset {\cal{C}}$ of cluster qubits measured
in measurement round $t$. Be $Q^{(t)} \subset {\cal{C}}$ the
set of qubits which are to be measured in the measurement round
$t$ and all subsequent rounds. Then, $Q_0$
is the set of qubits which are measured in the first round. These are
the qubits of which the 
observables $\sigma_x$, $\sigma_y$ or $\sigma_z$ are 
measured, so that the measurement bases are not influenced by other
measurement results. Further, $ Q^{(0)} = {\cal{C}}$. Now, the
sequence of sets $Q_t$ can be constructed using the following
recursion relation
\begin{equation}
    \label{Qrecur}
    \begin{array}{lcl}
        Q_t &=& \displaystyle{\left\{ q \in Q^{(t)} |
        \neg \exists p 
        \in {Q}^{(t)}: p \prec q \right\}} \\
        Q^{(t+1)} &=& \displaystyle{Q^{(t)} \backslash
        Q_t}.
    \end{array}
\end{equation}
All those qubits which have no precursors in some remaining set
$Q^{(t)}$ and thus do not have to wait for results of
measurements of qubits in $Q^{(t)}$ are taken out of this set
to form $Q_t$. The recursion proceeds until
${Q}^{(t_{\mbox{\footnotesize{max}}}+1)} = \emptyset$ for some maximal
value  $t_{\mbox{\footnotesize{max}}}$ of $t$.

Let us now discuss the classical processing. The scheme that emerges
is the following: 
The classical information gained by the measurements is
processed within a flow scheme. The flow quantity is a
classical $2n$-component binary vector ${\bf{I}}(t)$,
where $n$ is the number of logical qubits of a corresponding quantum  logic
network and $t$ the number of the measurement round. This vector
${\bf{I}}(t)$, the information flow vector, is 
updated after every measurement round. That is, after the one-qubit
measurements of all qubits of a set $Q_{t}$ have been performed
simultaneously, ${\bf{I}}(t-1$) is updated to
${\bf{I}}(t)$ through the results of these measurements. In
turn, ${\bf{I}}(t)$ determines which one-qubit observables are
to be measured of the qubits of the set $Q_{t+1}$. The result of the
computation is given by the information 
flow vector ${\bf{I}}(t_{\mbox{\footnotesize{max}}})$ 
after the last measurement round. From this
quantity the result of the readout measurement on the quantum register 
in the corresponding quantum logic network can be read off directly
without further processing.

In the following we briefly explain how this model arises. We
already mentioned the accumulated byproduct operator
$U_\Sigma|_\Omega$ which is the product of all the
$(U_{\Sigma,k})^{s_k}$, the forward propagated byproduct operators
randomly created by the measurement of qubits $k$ with outcome
$s_k$. $U_\Sigma|_\Omega$ determines how the readout has to be
interpreted. Now note that the readout measurement results can
themselves be expressed in terms of a byproduct operator. Let the
quantum register be in the state $|\psi_{\mbox{\footnotesize{out}}}\rangle=
\bigotimes_{i \in O} |s_i\rangle_{i,z}$ after readout. Then
$|\psi_{\mbox{\footnotesize{out}}}\rangle$ can be written as
$|\psi_{\mbox{\footnotesize{out}}}\rangle= U_R \, \bigotimes_{i \in O}
|0\rangle_i$ with $U_R =  \bigotimes_{i \in O}
(\sigma_x^{(i)})^{s_i}$, i.e. as a byproduct operator $U_R$ acting on
some standard state $|0\rangle_O$. This standard state contains no
information and can henceforth be discarded. The result of the
computation is contained in the byproduct operators.
It can be directly read off from the $x$-part of the operator
$U_{\Sigma,R}|_\Omega = U_\Sigma|_\Omega U_R$. 

If one discards the sign of these Pauli operators --an unphysical
global phase-- they can be mapped onto elements of a
$2n$-dimensional discrete vector space ${\cal{V}}$. For this we use the
isomorphism
\begin{equation}
    \label{Iso}   
        {\cal{I}}:\,\, I \in {\cal{V}}  \longrightarrow
         {\cal{P}}/\{\pm1\} \ni U =  \prod \limits_{i=1}^n
    \,\,{\left(\sigma_x^{(i)}\right)}^{{[I_x]}_i}
    {\left(\sigma_z^{(i)}\right)}^{{[I_z]}_i}, 
\end{equation}
where ${[I_x]}_i, {[I_z]}_i \in \{0,1\}$ are the respective components of
$I_x$, $I_z$ and ${\cal{P}}$ denotes the Pauli group. In particular,
the $x$-part of ${\bf{I}}={\cal{I}}^{-1} (U_{\Sigma,R}|_\Omega)$
represents the result of the quantum computation, corresponding to
the readout in the network model. 

To establish the terminology in which the classical information
processing with the \QC is described, we introduce the
{\em{byproduct images}} and the {\em{symplectic scalar product}}. The
byproduct image ${\bf{F}}_k$ of a cluster qubit $k \in {\cal{C}}$ is
defined as ${\bf{F}}_k={\cal{I}}^{-1}(U_{\Sigma,k}|_\Omega)$. Note that in
the byproduct operators for the implementation of the CNOT- and the
$\pi/2$-phase gate there are additional contributions which do not depend upon
measurement results, i.e. the byproduct operators are not the identity
for all measurement results being zero. These additional byproduct
operators have their byproduct images as well. Since they cannot be
related to a particular cluster qubit we attribute them to the gate
$g$ by whose implementation they are introduced and denote them by
${\bf{F}}_g$. 

Byproduct
images are easier to manipulate than the forward propagated byproduct
operators to which they correspond via
${\cal{I}}$. There hold the relations ${\cal{I}}({\bf{F}}_k + {\bf{F}}_l) =\pm
{\cal{I}}({\bf{F}}_k) {\cal{I}}({\bf{F}}_l)$, ${\cal{I}}(s_k
{\bf{F}}_k) = \pm{({\cal{I}}({\bf{F}}_k))}^{s_k}$. The symplectic
scalar product of two byproduct images ${\bf{F}}_k$, ${\bf{F}}_l$ is
defined as 
\begin{equation}
    \label{scalp}
    ({\bf{F}}_k,{\bf{F}}_l)_S = {\bf{F}}_{k,x}^T
    {\bf{F}}_{l,z} + {\bf{F}}_{k,z}^T {\bf{F}}_{l,x}\; \mbox{mod}\,2.
\end{equation} 
It is invariant under the Clifford group. A first application of the
objects introduced above is the cone test \cite{model},
\begin{equation}
    \label{conecrit}
    \forall\;k \in {\cal{C}},j\in Q^{(1)}: j\in \fc(k) \vee
    j\in \bc(k) \Leftrightarrow ({\bf{F}}_j,{\bf{F}}_k)_S=1. 
\end{equation}
Whether a qubit lies in some other qubits backward or forward
cone can be read off from the respective byproduct images.

We are now ready to discuss the process of computation. The goal is to
collect all the byproduct images weighted with the measurement
results, i.e. to finally obtain ${\bf{I}}={\cal{I}}^{-1}
(U_{\Sigma,R}|_\Omega)= \sum_g {\bf{F}}_g + \sum_{k \in {\cal{C}}} s_k
{\bf{F}}_k$ with all the cluster qubits measured in the correct basis. Before
the computation starts we propagate forward all the byproduct
operators attributed to the gates since they do not depend on any
measurement results. This reverses a number of angles specifying
one-qubit rotations in the network to simulate. In this way, we obtain
the algorithm angles
$\{\varphi^{\mbox{\footnotesize{init}}}_{j,\mbox{\footnotesize{algo}}}\}$
from the 
network Euler angles. Further we collect the byproduct images of the
gates, ${\bf{I}}_{\mbox{\footnotesize{init}}}=\sum_g {\bf{F}}_g$. 

In the first measurement round we measure all the cluster qubits $k
\in Q_0$ thereby removing the redundant qubits, implementing the
Clifford gates of the circuit and measuring the ``output
register''. This leaves us with byproduct operators scattered all over
the place. These byproduct operators we propagate forward and include
their byproduct images into the information flow vector at time $t=0$,
$
    {\bf{I}}(0) = {\bf{I}}_{\mbox{\footnotesize{init}}}+ \sum_{k \in
    Q_0} s_k {\bf{F}}_k.
$    
In forward propagation, the byproduct operators reverse some of the
algorithm angles. In this way, we update the algorithm angles
$\{\varphi^{\mbox{\footnotesize{init}}}_{j,\mbox{\footnotesize{algo}}}\}$
to the modified algorithm angles
$\{\varphi^{\mbox{\footnotesize{0}}}_{j,\mbox{\footnotesize{algo}}}\}$,
$\varphi^{\mbox{\footnotesize{0}}}_{j,\mbox{\footnotesize{algo}}}=
{(-1)}^{\eta_j}  
\varphi^{\mbox{\footnotesize{init}}}_{j,\mbox{\footnotesize{algo}}}$
with $\eta_j =\sum_{k \in Q_0\,|\, j \in \mbox{\footnotesize{fc}}(k)} s_k$.    
 
In subsequent measurement rounds we measure cluster qubits in adapted
bases. This also produces byproduct operators in the middle of the
network to simulate and we propagate them forward as before. The update of
${\bf{I}}(t)$ is just the same as in the first round. We find
\begin{equation}
    \label{Iup}
    {\bf{I}}(t)={\bf{I}}(t-1) + \sum_{k \in Q_t} s_k {\bf{F}}_k =
    {\bf{I}}_{\mbox{\footnotesize{init}}}+ \sum_{k \in 
    \bigcup_{r=0}^t Q_r} s_k {\bf{F}}_k. 
\end{equation}
After the final update, ${\bf{I}}(t_{\mbox{\footnotesize{max}}}) =
{\bf{I}}$ contains the result of the computation in its $x$-part.

As in the first round, the propagation of byproduct operators affects
the angles 
that specify the measurement bases. The update of these angles in all the
subsequent measurement rounds could be performed in the same way as in 
the first round providing one with a complete history
$\{\varphi^{\mbox{\footnotesize{t}}}_{j,\mbox{\footnotesize{algo}}}\}$
of adapted angles. It is not necessary to generate and store this bulk
of information. We only need to know the adapted angles
$\varphi_{j,\mbox{\footnotesize{meas}}}$ at the time when the
respective qubits are measured. These measurement angles
$\varphi_{j,\mbox{\footnotesize{meas}}}$ can be obtained in a more
compact procedure.  

This leads to the question which measurement outcomes affect the
choice of the measurement basis at qubit $j \in Q_t,\,\, t>0$. All the
measurement outcomes obtained at qubits 
$k \in {\cal{C}}$ with $j \in \mbox{fc}(k)$ contribute, i.e.
\begin{equation}
    \label{phimeas}
    \begin{array}{rcl}
        \varphi_{j,\mbox{\footnotesize{meas}}} &=&
        \varphi^{\mbox{\footnotesize{init}}}_{j,\mbox{\footnotesize{algo}}}\,
        (-1)^{\sum_{k \in
        {\cal{C}}\,|\, j \in \mbox{\tiny{\fc}}(k)} s_k} \\
        &=& \varphi^{\mbox{\footnotesize{0}}}_{j,\mbox{\footnotesize{algo}}}
        \, (-1)^{\sum_{k \in
        {\cal{C}}\backslash Q_0\,|\, j \in \mbox{\tiny{\fc}}(k)} s_k}.
    \end{array}
\end{equation}  
In the second line of (\ref{phimeas}) we can now simplify the sign
factor by use of the cone test (\ref{conecrit}). First note that the
sum over $k \in {\cal{C}}\backslash Q_0$ reduces to a sum over $k \in
\bigcup_{r=1}^{t-1} Q_r$ because otherwise $j \not\in
\mbox{fc}(k)$. Then, for $k \in Q_r, j \in
Q_t$, $ 1 < r < t$ qubit $j$ may only be in the forward cone of qubit
$k$, but never in the backward cone $\mbox{bc}(k)$. Hence, the cone
test simplifies to $j\in \fc(k) \Longleftrightarrow
({\bf{F}}_j,{\bf{F}}_k)_S=1$ for such qubits, and we obtain 
$
\sum_{k \in {\cal{C}}\backslash Q_0\,|\, j \in \mbox{\tiny{\fc}}(k)}
s_k = \sum_{k \in \bigcup_{r=1}^{t-1} Q_r} s_k ({\bf{F}}_k,{\bf{F}}_j)_S =
({\bf{I}}(t-1)-{\bf{I}}(0),{\bf{F}}_j)_S
$, and thus
\begin{equation}
    \label{phimeas2}
   \varphi_{j,\mbox{\footnotesize{meas}}} =
   \varphi^\prime_{j,\mbox{\footnotesize{algo}}}\, (-1)^{({\bf{I}}(t-1),
   {\bf{F}}_j)_S},  
\end{equation}
with $\varphi^\prime_{j,\mbox{\footnotesize{algo}}} =
\varphi^0_{j,\mbox{\footnotesize{algo}}}\, (-1)^{({\bf{I}}(0), 
  {\bf{F}}_j)_S}$. If one works this out one finds that the angles
$\varphi^\prime_{j,\mbox{\footnotesize{algo}}}$ are obtained from
the corresponding Euler angles of the network by 
propagating the byproduct operators of the gates and of the qubits
measured in the first round {\em{backwards}} to the input side of the network. 

To sum up, the $2n$-component binary valued information flow vector
represents the information that is processed with the \QCns. Although
random in its numerical value after all measurement rounds but the
final one, it 
has a meaning in every step of the computation. The rule for the adaption
of measurement bases (\ref{phimeas2}) invokes the random measurement results of
qubits $k \in {\cal{C}}\backslash Q_0$ 
{\em{only}} via the information flow vector. The measurement results on
qubits $k \in Q_0$ are absorbed into the angles
$\varphi^\prime_{j,\mbox{\footnotesize{algo}}}$ and can be erased
after these angles have been set. The angles
$\varphi^\prime_{j,\mbox{\footnotesize{algo}}}$ remain unchanged in
the further course of computation. After the
final update at $t=t_{\mbox{\footnotesize{max}}}$, when there are no
measurement bases left to adjust, the information flow vector
${\bf{I}}(t_{\mbox{\footnotesize{max}}})$ displays the result of the
computation. The update of ${\bf{I}}(t)$ (\ref{Iup}) and the rule to
adjust the measurement angles (\ref{phimeas2}) are very simple
algebraic operations.

\section{Conclusion}

We have reviewed the computational model underlying the
one-way quantum computer \cite{model}, which is very different from the
quantum logic network model. The logical depth of
certain algorithms is on the \QC lower than has so far been known for
networks.  
As an example, on the \QC circuits composed of CNOT-,
Hadamard- and $\pi/2$-phase gates have unit logical depth, independent
of the number of gates or logical qubits. The best bound for   
networks known previously scales logarithmically. It
therefore seems that the question of temporal complexity must be revisited.

The formal description of the \QC is based on primitive
quantities of which the most important are the sets $Q_t \subset
{\cal{C}}$ of cluster 
qubits defining the temporal ordering of measurements on the cluster
state, and the binary valued information flow vector ${\bf{I}}(t)$
which is the carrier of the algorithmic information. Much of the
terminology that one is familiar with from the 
network model has been abandoned since in case of the \QC no proper meaning
can be assigned to these objects. In fact, the \QC has no quantum input, no
quantum output, no quantum register and it does not consist of quantum gates.

The \QC is nevertheless quantum mechanical as it uses a highly
entangled cluster state as the central physical resource. It works by measuring
quantum correlations of the universal cluster state.

\section*{Acknowledgements}

This work has been supported by the Deutsche Forschungsgemeinschaft
(DFG) within the Schwerpunktprogramm QIV. We would like to thank D. E. 
Browne and H. Wagner for helpful discussions.

\end{document}